\documentclass[11pt]{article}
\usepackage{moriond,epsfig,graphics}

\def\pio{\pi^0}

\def\be{\begin{equation}}
\def\ee{\end{equation}}
\def\bea{\begin{eqnarray}}
\def\eea{\end{eqnarray}}

\begin{document}
\vspace*{4cm}
\title{DI-PHOTON AND PHOTON+B/C PRODUCTION CROSS SECTIONS AT ECM=2TEV}

\author{ R. MCNULTY\\(on behalf of the CDF collaboration) }

\address{Department of Physics, University College Dublin, \\
Belfield, Dublin 4, Ireland.}

\maketitle\abstracts{
The inclusive di-photon cross-section is measured in the central region and found to be in good agreement
with NLO QCD predictions.  Cross-sections are presented for events containing
an energetic photon in addition to a heavy flavour jet.  The ratio of photon+c
to photon+b events is also measured.  Results are currently statistically limited and in
agreement with Pythia predictions.
}

\section{Introduction}
Events containing photons can be used to test QCD predictions and to explore physics processes beyond the Standard Model.  
We describe the results of two analyses using data taken by the CDF collaboration: one measures the cross-section for inclusive 
di-photon production; the other the cross-section for events containing photons accompanied by heavy flavour jets.

\section{Photon Detection at CDF}
RunII at the Tevatron collider started in 2001 and by the end of 2003, the CDF experiment had recorded over 300$pb^{-1}$ of data.  
Between $67pb^{-1}$ and $207pb^{-1}$ have been used in the analyses presented here.  Photons are detected in CDF as the presence of 
deposits in the electromagnetic calorimeters which are not associated with charged track extrapolations.  The main background 
to the identification of single photons comes from the decays of neutral mesons and in particular energetic $\pio$s.  
Two independent methods are 
used in order to separate the signal from background.

\subsection{The CES method}
The first method depends on the Central Electromagnetic Shower (CES) profile.  Energetic $\pio$s decay to two overlapping photons 
which can not be resolved in the calorimeter.  The CES, situated at the position of maximum lateral shower extent inside the
calorimeter, has finer segmentation and can distinguish between photons and $\pio$s on the basis of shower profile. 
The CES method uses this information to assign a
weight to each electromagnetic deposit according to the likelihood of it being due to a photon or a $\pio$.  This method is most 
effective for lower energy $\pio$s where the photons are more separated; above 35GeV the opening angle of the two photons is too 
small to allow a discrimination.

\subsection{The CPR method}
The second method makes uses of the Central PreRadiator (CPR) which is a 
gas chamber located on the front face of the central calorimeter.
The probability for a photon to convert before reaching the CPR is about 60\%.  
A signal in the CPR is thus more likely to originate from a $\pio$ than a single photon
and the response of the CPR can be used to assign a likelihood to a calorimeter deposit being
due to a single photon.
This method is applicable at all photon energies and is used in particular above 35GeV.
 
\section{Measurement of the di-photon Cross-section}
The motivations for a measurement of the cross-section of  events containing 
two isolated photons are threefold: firstly, it is 
a test of QCD;  secondly it has sensitivity to new physics processes;  
and thirdly it will be the major background process to a Higgs search 
at the LHC~\cite{higgs} for a Higgs with mass below 150GeV, where the most promising detection mechanism
appears to be $H\rightarrow\gamma\gamma$.

The analysis requires two energetic photons, 
one greater than 14GeV the other above 13GeV, within a rapidity of $|\eta|<0.9.$.
From $207pb^{-1}$ of 
data a sample of 573 candidates was obtained of which 45\%
were estimated to be di-photon events.
The distribution of events was compared to three simulations: ResBos~\cite{ResBos};
DIPHOX~\cite{diphox}; and Pythia~\cite{pythia}.
Both ResBos and DIPHOX are NLO predictions whilst Pythia is a LO calculation.  
DIPHOX includes NLO fragmentation contributions while ResBos 
uses LO fragmentation.~\cite{ResBoslo}

\begin{figure}
\mbox{\psfig{figure=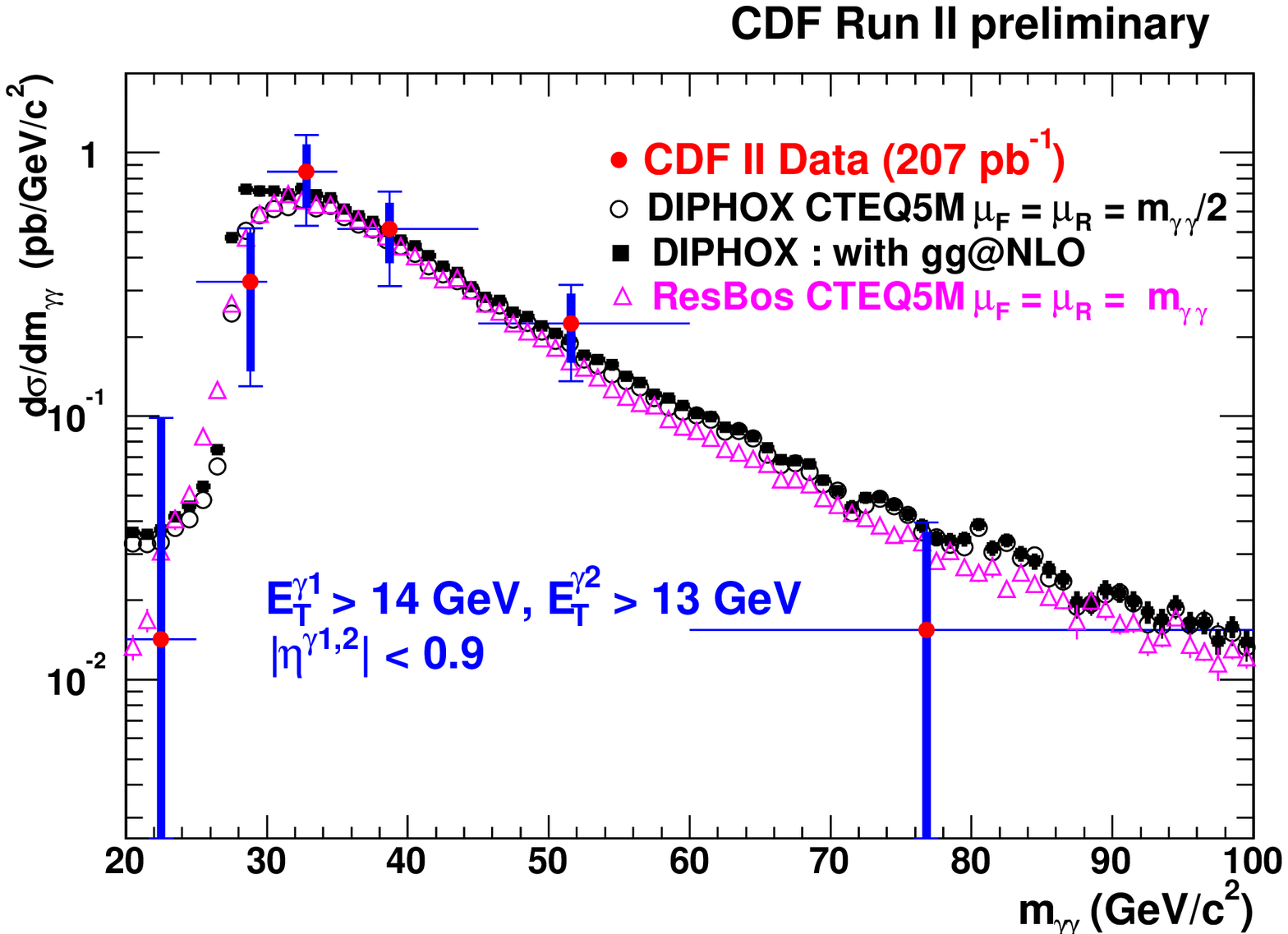,height=2.2in}
\psfig{figure=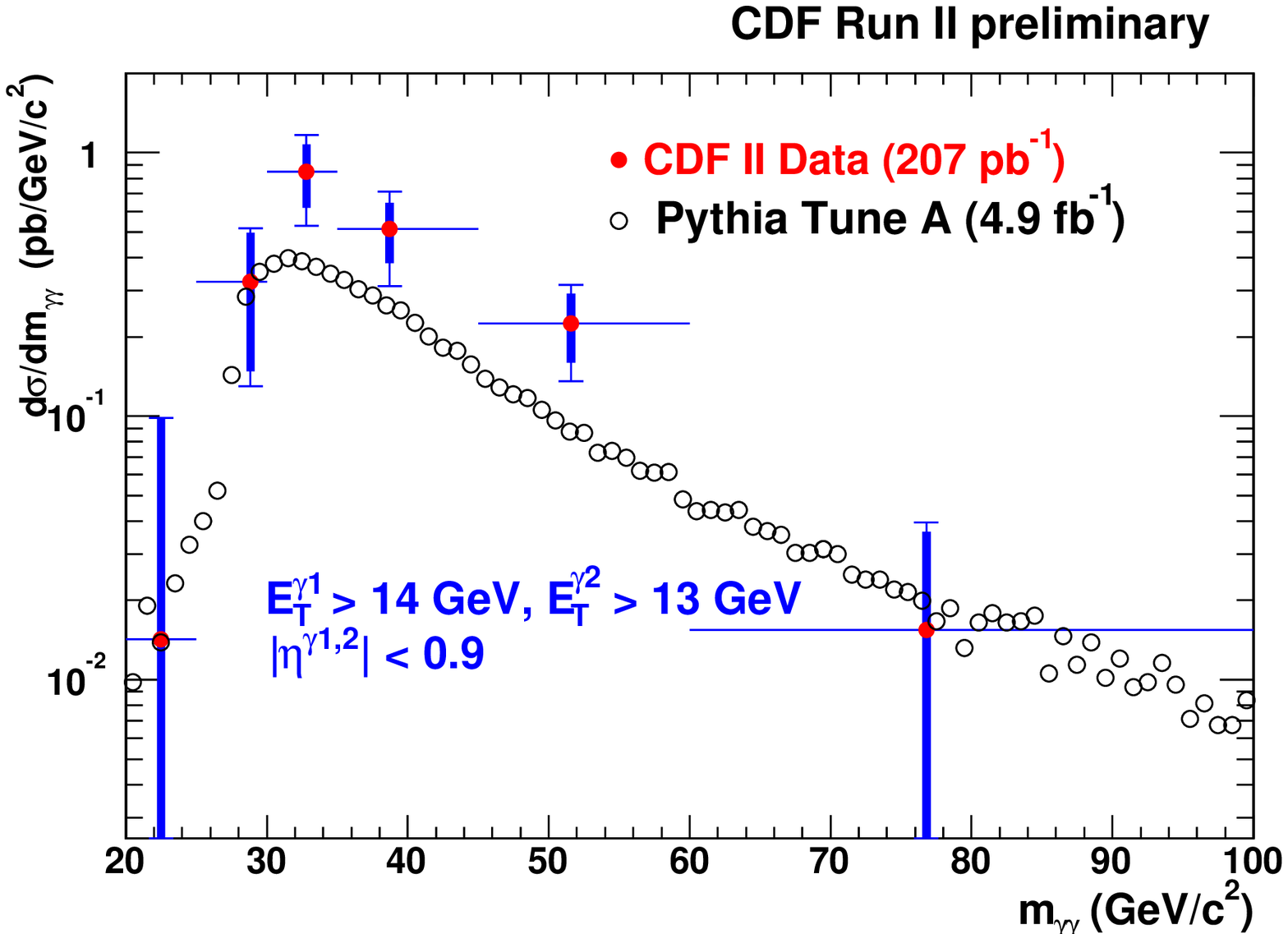,height=2.2in}}
\caption{Invariant mass of di-photons for data compared to DIPHOX and ResBos (left) and Pythia (right).
\label{fig:mm}}
\end{figure}

Fig.~\ref{fig:mm} shows the differential cross-section as a function of the invariant mass of the di-photon system.  
The data are compared to DIPHOX 
and ResBos, both using CTEQ5M structure functions.  
Good agreement is observed between the data and both simulations.  
The dominant production process for events producing a low invariant mass of the 
two photons is through gluon-gluon fusion while high invariant mass events are principally produced by 
quark annihilation.
DIPHOX 
originally included just a LO contribution for the gluon-gluon contribution.  
The effect of adding higher order corrections for 
gluon-gluon contribution~\cite{diphoxhi} results in a slight change to the spectrum at low mass, as expected.  However, the 
data is not yet sensitive enough to discriminate between these predictions.  
On the other hand, as also shown in Fig.\ref{fig:mm}, the data is not 
in agreement with (leading order) Pythia, 
the data points lying significantly above the 
theoretical curve; nonetheless the shape (after tuning~\cite{rick}) is consistent.

\begin{figure}
\hspace{4cm}
\psfig{figure=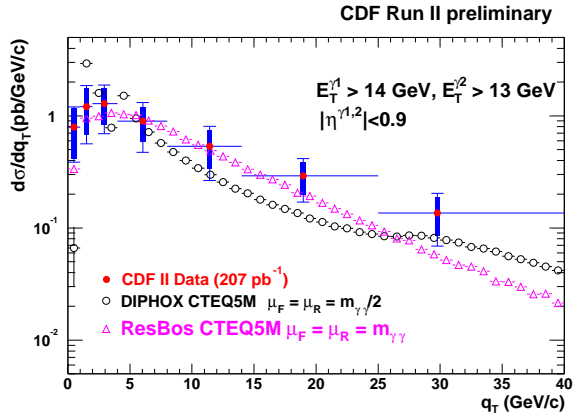,height=2.2in}
\caption{Distribution of $q_T$ of the di-photon system for data compared to DIPHOX and ResBos.
\label{fig:qt}}
\end{figure}

Perhaps more interesting from a theoretical point of view and to make predictions at LHC energies is the differential cross-section as a 
function of $q_T$, the transverse momentum of the photon system.  
Fig.~\ref{fig:qt} shows the data compared to both DIPHOX and ResBos.  In common 
to any fixed order calculation, DIPHOX suffers infra-red divergences.  
Due to the isolation requirements imposed experimentally to detect
photons, this results in a singularity 
(see \cite{diphox} for more discussion) 
the effect of which is visible at about 4GeV where 
the prediction is untrustworthy.  In contrast, ResBos resums the effects of soft and/or collinear gluon emissions to all orders and predicts 
a smooth $q_T$ distribution.  

\section{Measurements of Photon+b and Photon+c Cross-sections}
These measurements test QCD predictions of heavy flavour production and 
are sensitive to physics beyond the standard model, for example 
the production of excited quarks or GMSB where neutralinos decay radiatively to gravitinos~\cite{susysearch}.
A measurement of the ratio of both processes is also of interest.
It might be expected that photon+c events will be four times more plentiful than photon+b events.  However this prediction 
must be modified by the relative number of b and c quarks events.  In particular, the theoretical prediction includes a contribution coming from the 
charm content of the proton and thus the experimental measurement 
has sensitivity to this structure function.

The analysis chose events with a photon of $|\eta|<1$ and energy above 25GeV  accompanied by a jet within which a secondary vertex could be 
determined.  Secondary vertices could be due to the lifetime content of b- or c-quark events or sometimes to the combination of a poorly 
measured track with other tracks.  The relative contributions coming from events containing b,c, and other quarks were determined from the 
distribution of the invariant mass of tracks associated to the secondary vertex.  Simulations showed that b-induced events had higher average 
masses than c-induced events which in turn were more massive than other events.  Templates representative of each species
were fit to the data spectrum in order to calculate the relative amounts of each contribution.

The fits were performed in four bins of photon transverse energy: 25-29GeV; 29-34GeV; 34-42GeV; and 42-60GeV.
The fake photon background was calculated from data using the CPR method.  The estimates of the number of photon+b 
and photon+c events in the sample were turned into a cross-section on dividing by the luminosity and the efficiencies 
for observing a photon and for tagging a secondary vertex.

\begin{figure}

\rotatebox{270}{\resizebox{0.32\textwidth}{!}{\includegraphics{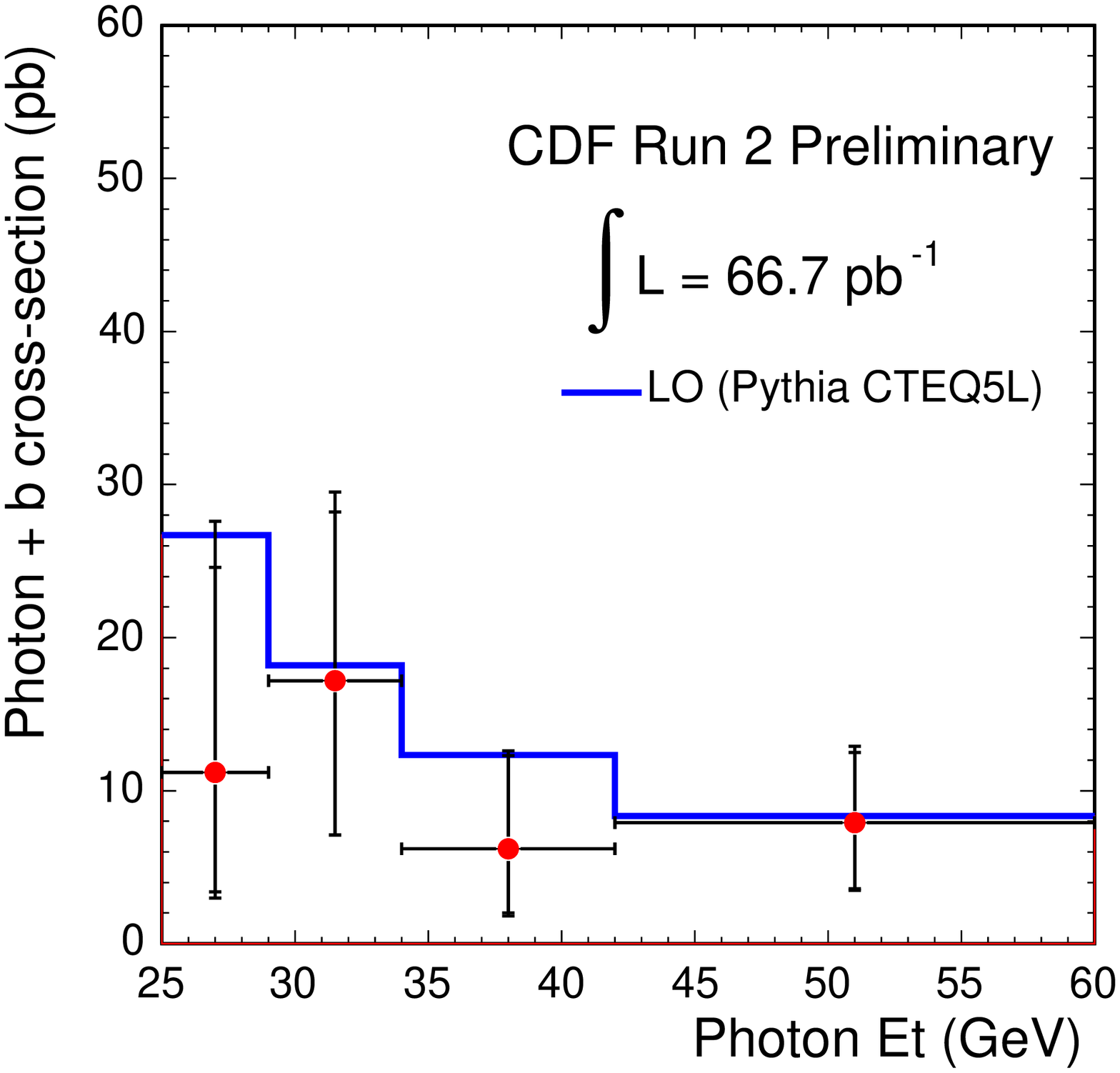}}}
\hspace{1.4cm}
\rotatebox{270}{\resizebox{0.32\textwidth}{!}{\includegraphics{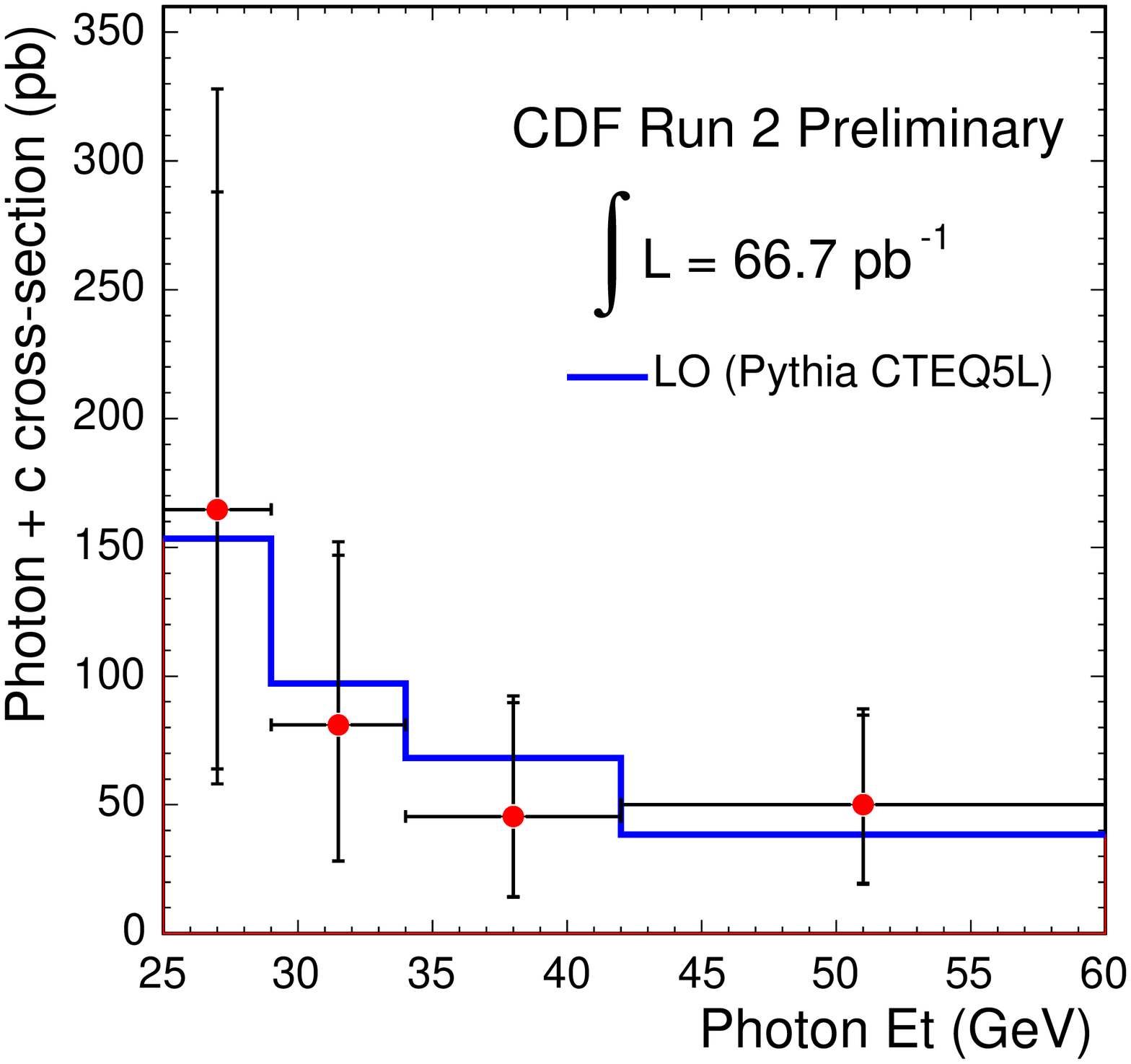}}}
\hspace{1.4cm}
\rotatebox{270}{\resizebox{0.32\textwidth}{!}{\includegraphics{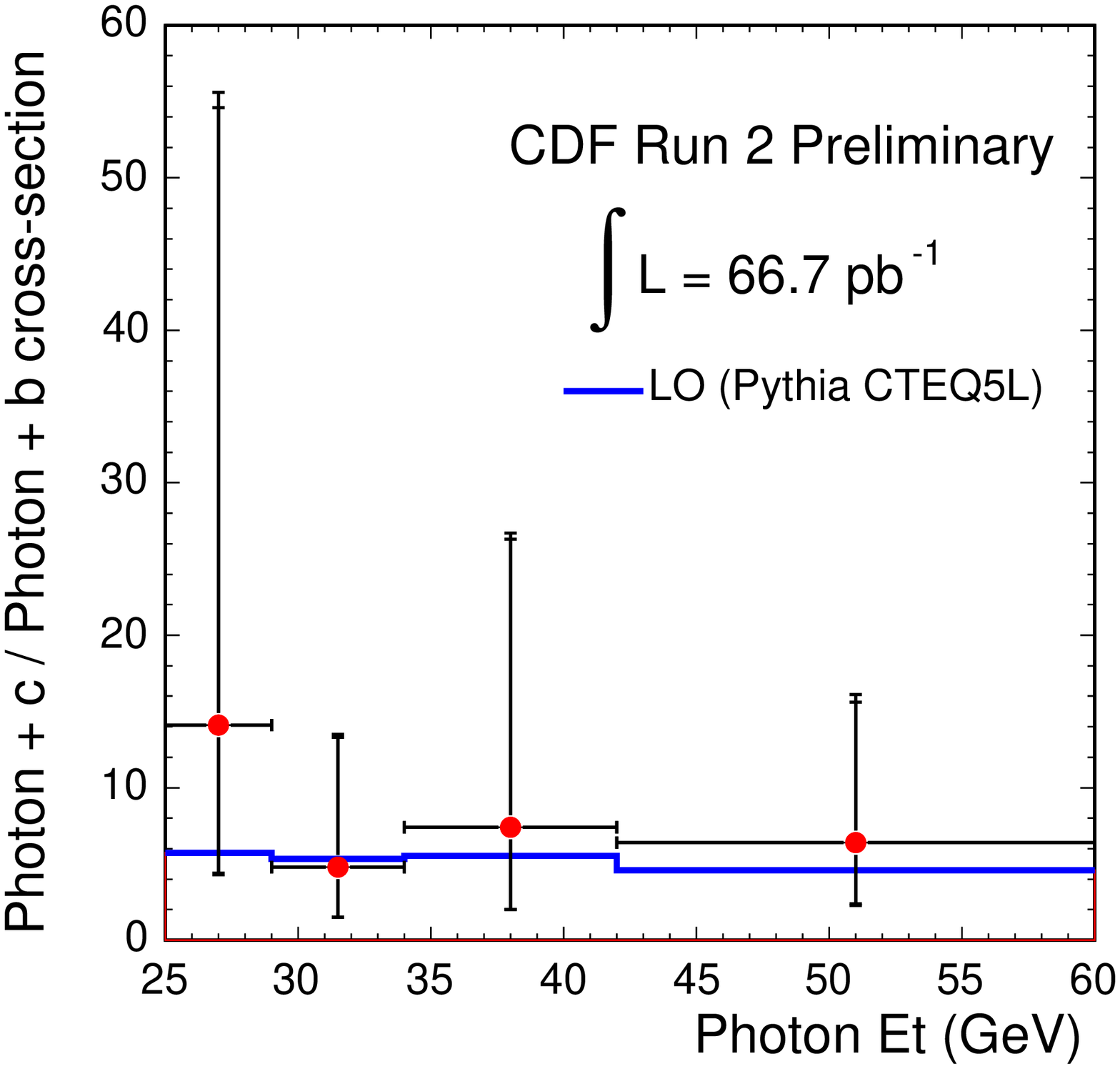}}}

\caption{Cross-sections for events containing photons plus b-quarks (left) 
and photons plus c-quarks (centre).  Points are data; the line is the Pythia prediction.
The ratio of photon+c to photon+b is plotted on the right.
\label{fig:bresult}}
\end{figure}

The results are shown in Fig.~\ref{fig:bresult} and are seen to be consistent with the leading order Pythia prediction.
The measurements are clearly statistically limited at present.  The dominant systematics come from the estimation of the 
heavy-flavour tagging 
efficiency and from the determination of the jet energy scale.  The latter, in addition to other systematics are 
significantly reduced by instead 
considering the ratio of cross-section for photon+c to photon+b.  

\section{Conclusions}

Measurements have been made of the cross-sections in the central region for inclusive di-photon production and for photons accompanied by 
b-quarks and c-quarks.  With $207pb^{-1}$ the di-photon cross-section is in agreement with NLO predictions of DIPHOX and ResBos and disagrees 
with LO Pythia.  With 
$67pb^{-1}$, the photon + heavy flavour production agrees with Pythia LO predictions although the measurement is currently statistically limited.  
The Tevatron continues to produce data which will allow updates and improvements to both these analyses in the near future.


\section*{References}
\begin{thebibliography}{99}
\bibitem{higgs} ATLAS Collaboration, Detector and Physics Performance TDR, CERN/LHCC/99-14;\\
CMS Collaboration, Electromagnetic Calorimeter TDR, CERN/LHCC/97-33.
\bibitem{ResBos} C.\ Balazs et al., {\it Phys. Rev.}{\bf D 57} (1998) 6934.
\bibitem{diphox} T. Binoth et al., {\it Eur. Phys. J.} {\bf C 16} (2000) 311.
\bibitem{pythia} T.\ Sjöstrand et al. {\it Computer Phys. Commun.} {\bf 135} (2001) 238.
\bibitem{ResBoslo}T.\ Binoth et al.,{\it Phys.Rev.} {\bf D63} (2001)114016.  
\bibitem{diphoxhi} Z.\ Bern et al., {\it Phys.Rev.}{\bf D66}(2002) 074018.
\bibitem{rick} T.\ Affolder et al.,{\it Phys. Rev.}{\bf D65} (2002) 092002;\\
See also {\tt http://www.phys.ufl.edu/~rfield/cdf/tunes/rdf\_tunes.html}
\bibitem{susysearch} T.\ Affolder et al.,{\it Phys. Rev.}{\bf D 65} (2002) 052006.

\end{thebibliography}
\end{document}